\documentclass[twocolumn]{aastex631}

\newcommand\HI{H\,{\sc i}}
\newcommand\Htwo{H$_2$}
\newcommand\SHI{$\Sigma_{\rm H\,{\footnotesize{\textsc i}}}$}
\newcommand\SHtwo{$\Sigma_{\rm H_2}$}

\newcommand\Ss{$\Sigma_*$}

\usepackage{soul}

\begin{document}

\title{VERTICO and IllustrisTNG: The spatially resolved effects of environment on galactic gas}

\author[0000-0003-1908-2168]{Adam R.~H.~Stevens}
\altaffiliation{adam@a4e.org}
\affiliation{International Centre for Radio Astronomy Research, The University of Western Australia, Crawley, WA 6009, Australia}

\author[0000-0003-1845-0934]{Toby Brown}
\altaffiliation{tobias.brown@nrc-cnrc.gc.ca}
\affiliation{Herzberg Astronomy and Astrophysics Research Centre, National Research Council of Canada, Victoria, BC, V9E 2E7, Canada}


\author[0000-0001-9568-7287]{Benedikt Diemer}
\affiliation{Department of Astronomy, University of Maryland, College Park, MD 20742, USA}

\author[0000-0003-1065-9274]{Annalisa Pillepich}
\affiliation{Max-Planck-Institut f\"{u}r Astronomie, D-69117 Heidelberg, Baden-W\'{u}rttemburg, Germany}

\author{Lars Hernquist}
\affiliation{Institute for Theory and Computation, Harvard--Smithsonian Center for Astrophysics, Cambridge, MA 02138, USA}

\author[0000-0001-8421-5890]{Dylan Nelson}
\affiliation{Institut f\"{u}r theoretische Astrophysik, Zentrum f\"{u}r Astronomie, Universit\"{a}t Heidelberg, D-69120 Heidelberg, Baden-W\"{u}rttemburg, Germany}


\author{Yannick M.~Bah\'{e}}
\affiliation{Institute of Physics, Laboratory of Astrophysics, Ecole Polytechnique F\'{e}d\'{e}rale de Lausanne (EPFL), Observatoire de Sauverny, 1290 Versoix, Switzerland}
\affiliation{Leiden Observatory, Leiden University, PO Box 9513, NL-2300 RA Leiden, The Netherlands}

\author[0000-0002-9795-6433]{Alessandro Boselli}
\affiliation{Aix-Marseille Universit\'{e}, CNRS, CNES, LAM, Marseille, France}

\author[0000-0003-4932-9379]{Timothy A. Davis}
\affiliation{Cardiff Hub for Astrophysics Research \& Technology, School of Physics \& Astronomy, Cardiff University, Queens Buildings, Cardiff, CF24 3AA, UK}

\author[0000-0002-6154-7224]{Pascal J. Elahi}
\affiliation{Pawsey Supercomputing Centre, Kensington, WA 6151, Australia}

\author[0000-0002-1768-1899]{Sara L.~Ellison}
\affiliation{Department of Physics \& Astronomy, University of Victoria, Finnerty Road, Victoria, BC V8P 1A1, Canada}

\author[0000-0002-9165-8080]{Mar\'{i}a J.~Jim\'{e}nez-Donaire}
\affiliation{Department of Astronomy, University of Maryland, College Park, MD 20742, USA}
\affiliation{Observatorio Astron\'{o}mico Nacional (IGN), C/Alfonso XII, 3, E-28014 Madrid, Spain}

\author[0000-0002-0692-0911]{Ian D.~Roberts}
\affiliation{Leiden Observatory, Leiden University, PO Box 9513, NL-2300 RA Leiden, The Netherlands}

\author[0000-0002-0956-7949]{Kristine Spekkens}
\affiliation{Royal Military College of Canada, PO Box 17000, Station Forces, Kingston, ON, K7K 7B4, Canada}

\author[0000-0002-5877-379X]{Vicente Villanueva}
\affiliation{Department of Astronomy, University of Maryland, College Park, MD 20742, USA}

\author[0000-0002-9405-0687]{Adam B.~Watts}
\affiliation{International Centre for Radio Astronomy Research, The University of Western Australia, Crawley, WA 6009, Australia}

\author[0000-0001-5817-0991]{Christine D.~Wilson}
\affiliation{Department of Physics \& Astronomy, McMaster University, 1280 Main Street W, Hamilton, ON, L8S 4M1, Canada}

\author[0000-0001-7732-5338]{Nikki Zabel}
\affiliation{Department of Astronomy, University of Cape Town, Private Bag X3, Rondebosch 7701, South Africa}

\begin{abstract}
\noindent
It has been shown in previous publications that the TNG100 simulation quantitatively reproduces the observed reduction in each of the total atomic and total molecular hydrogen gas for galaxies within massive halos, i.e.~dense environments.
In this Letter, we study how well TNG50 reproduces the \emph{resolved} effects of a Virgo-like cluster environment on the gas surface densities of satellite galaxies with $m_* > \! 10^9\,{\rm M}_\odot$ and ${\rm SFR} \! > 0.05\,{\rm M}_\odot\,{\rm yr}^{-1}$.
We select galaxies in the simulation that are analogous to those in the HERACLES and VERTICO surveys, and mock-observe them to the common specifications of the data.
Although TNG50 does not quantitatively match the observed gas surface densities in the centers of galaxies, the simulation does qualitatively reproduce the trends of gas truncation and central density suppression seen in VERTICO in both \HI~and \Htwo.
This result promises that modern cosmological hydrodynamic simulations can be used to reliably model the post-infall histories of cluster satellite galaxies.
\end{abstract}



\section{Introduction}
\label{sec:intro}
In recent years, the realism of cosmological hydrodynamic simulations has grown to the point where the empirical effects of galaxy environment on the global gas properties of galaxies are demonstrably reproducible at low redshift
\citep[e.g.][]{stevens19a,stevens21,yun19}.
Indeed, this is also true for some semi-analytic models of galaxy formation \citep[e.g.][]{sb17,xie20}.
These tests of forefront models in the literature have become possible thanks to statistically significant and representative observational surveys that have traced the emission from \emph{both} the atomic and molecular gas in galaxies that cover the spectrum of environments from field isolation to massive galaxy clusters at low redshift \citep[e.g.][]{catinella18}.

While the broad effects of environment on galaxies' gas is now reasonably well understood \citep[see the review by][]{cortese21}, surveys have started directing their attention to the gas properties on local scales inside galaxy discs.
The `Virgo Environment Traced In CO' survey \citep[VERTICO,][]{brown21} has observed the molecular hydrogen gas content [\Htwo, traced via the CO(2--1) line] at sub-kpc resolution across 51 late-type galaxies in the Virgo cluster, all of them with existing multi-wavelength maps of atomic hydrogen gas (\HI), stellar emission, and star formation activity. 
Recently, \citet{watts23} used this dataset to show that, as galaxies are processed by the cluster, environmental mechanisms drive a continuous decrease in both \HI~and \Htwo~local surface densities (\SHI~and \SHtwo, respectively) at fixed local stellar surface density (\Ss) with respect to the field. 
From a theoretical perspective, it is clearly an important next step to establish if the latest simulations are capable of reproducing (i) the scaling relations of \SHI~and \SHtwo~values with \Ss~observed in nearby field and cluster populations, and (ii) the systematic influence of environment upon those relationships.

Taking that step in this Letter,
we investigate the resolved gas scaling relations of individual galaxies from the TNG50 simulation \citep{nelson19,pillepich19}, mock-observing them to be directly comparable to VERTICO and the `Heterodyne Receiver Array CO Line Extragalactic Survey' \citep[HERACLES;][]{leroy09}.


\section{Data and methods}
\label{sec:data}

\subsection{Observations}
\label{ssec:obs}

The observational samples of field and cluster galaxies used in this paper are drawn from HERACLES \citep{leroy09} and VERTICO \citep{brown21}, respectively. The analysis sample used in this paper is a super-set of that used in \citet{watts23}, and there is a detailed discussion of the observational biases present in comparing these data in that work. Briefly, the two data sets contain star-forming late-type galaxies that are well matched in global stellar mass and specific SFR. The \HI observations for VERTICO and HERACLES galaxies are drawn from the VLA Imaging of Virgo in Atomic gas survey \citep[VIVA;][]{chung09} and  The HI Nearby Galaxy Survey \citep[THINGS;][]{walter08}, respectively. All molecular gas information is derived from the public data cubes using the methodology described in \citet{brown21}. Similarly, global stellar-mass and SFR estimates are drawn from the $z \! = \! 0$ Multi-wavelength Galaxy Synthesis database \citep[$z0$MGS;][]{leroy19}, while resolved stellar and SFR surface densities are derived from identical multiwavelength data for both surveys following the methods outlined in \citet{villanueva22} and \citet{jimenez23}, respectively.
VERTICO is typically 2--3 times more sensitive than HERACLES in both \HI~and \Htwo~densities, which ensures that any observed differences with environment are not driven by sensitivity differences between the surveys.

For this work, we select we only select galaxies with stellar mass $m_* \! > \! 10^9\,{\rm M}_\odot$ and inclinations less than 70$^\circ$, ensuring the galaxies' surface density maps can be reliably deprojected with a simple cosine-of-inclination correction factor. 
These selections yield a final resolution-matched sample consisting of 10 field galaxies from HERACLES and 33 cluster galaxies from VERTICO.
These galaxies all have star formation rates ${\rm SFR} \! \gtrsim \! 0.05\,{\rm M}_\odot\,{\rm yr}^{-1}$.

Very briefly, each galaxy has molecular gas, stellar mass, and SFR surface density maps that have been smoothed to a spatial resolution of $\sim$1.2\,kpc with a pixel size of $\sim$650\,pc to approximately Nyquist-sample the smoothing kernel [or CO(2--1) resolving beam]. 
This approximately matches the resolution of the \HI~data from VIVA (including an update where the D-configuration data were removed, leaving only the higher-resolution C-configuration data).
The derived data products for VERTICO and HERACLES are produced using a near identical procedure to that described in \citet{brown21} for the molecular gas surface densities, \citet{jimenez23} for the star formation rates, and \citet{watts23} for the stellar surface densities.
The only difference between the data presented in previous VERTICO papers and here is that we have reduced the molecular gas surface densities by a factor of 1.36 to remove the `helium contribution', as we are specifically interested in \Htwo.


\subsection{Simulations}
\label{ssec:sims}
The IllustrisTNG model of galaxy formation \citep{weinberger17,pillepich18a} comprises key descriptions of astrophysical processes, including gas cooling, star and black-hole formation, stellar evolution and feedback, feedback from active galactic nuclei (AGN), and more, all within a $\Lambda$CDM cosmological, magnetohydrodynamic framework with the moving-mesh {\sc arepo} code \citep{springel10}.
Simulations of various box sizes and resolutions have been run with the IllustrisTNG model, all of which carry identical parameters for both the sub-resolution physical prescriptions and cosmology, with the latter based on \citet{planck16}.
We use the TNG50 simulation \citep{nelson19,pillepich19} in this paper, which has a periodic box length of $\sim$50\,cMpc and baryonic mass resolution of $8.5 \!\times\! 10^4\,{\rm M}_\odot$.
Galaxy subhalos in TNG are identified with {\sc subfind} \citep{springel01,dolag09}.

The \HI~and \Htwo~properties of gas cells in TNG were calculated in post-processing \citep{diemer18,stevens19a}.
In this paper, we use the decomposition based on \citet{gd14} as described in \citet{stevens19a}.

\begin{figure}[t]
\centering
\includegraphics[width=0.98\columnwidth]{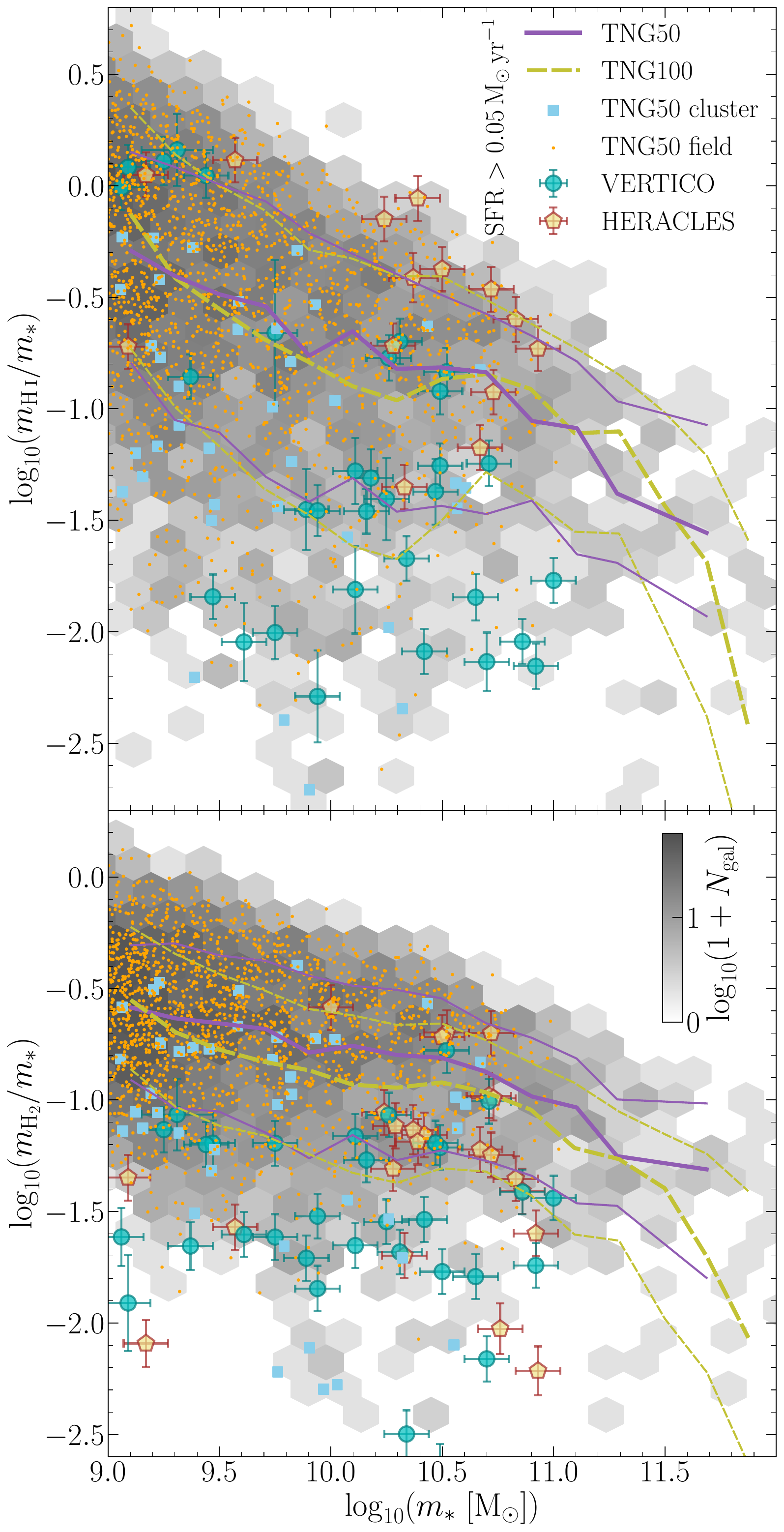}
\vspace*{-1.0cm}\caption{The \HI~(top panel) and \Htwo~(bottom panel) fractions of TNG50 and TNG100 galaxies as a function of stellar mass at $z\!=\!0$.
Only galaxies with $m_* \! \geq \! 10^9\,{\rm M}_\odot$ and ${\rm SFR} \! \geq 0.05\,{\rm M_\odot\,yr}^{-1}$ are included (without any environmental sub-sampling).
Hex bins show the number density of TNG50 galaxies.
Lines are the running median (thick) and 16th and 84th percentiles (thin) for TNG50 (solid) and TNG100 (dashed).
Points with approximate errors compare the VERTICO and HERACLES galaxies that we use in this paper (a subset of the full surveys; cf.~fig.~1 of \citealt{zabel22}).
We show these observational data for reference, but we do \emph{not} necessarily expect the simulation medians to align closely with them (but they should be closer to HERACLES than VERTICO).
Individual points from our TNG50 cluster and field samples, shown, can be respectively compared to VERTICO and HERACLES.}

\label{fig:HIH2frac}
\end{figure}

The integrated \HI~and \Htwo~properties of TNG100 galaxies and their trends with environment have been explored in depth \citep{diemer19,stevens19a,stevens21}, but the same is not true in the literature%
\footnote{Figures published in \citet{diemer19} with TNG100 and TNG300 have been reproduced with TNG50 at \url{http://www.benediktdiemer.com/data/hi-h2-in-illustris/}.}
for TNG50 \citep[but see][]{boselli23}.
In brief, after mock-observing the simulated galaxies to match survey specifications, \citet{stevens19a,stevens21} show that TNG100 quantitatively reproduces gas-fraction trends seen with ALFALFA, xGASS, and xCOLD GASS \citep[i.e.~from the results of][]{brown17,saintonge17,catinella18}.
To show that the gas fractions of TNG50 and TNG100 are consistent, we compare the \HI~and \Htwo~fractions of the simulations in Fig.~\ref{fig:HIH2frac}.
To approximately match the galaxies in the observational samples described in Section \ref{ssec:obs}, we exclusively consider TNG galaxies at $z\!=\!0$ with $m_* \! \geq \! 10^9\,{\rm M}_\odot$ and ${\rm SFR} \! \geq 0.05\,{\rm M_\odot\,yr}^{-1}$ (based on measurements internal to the `BaryMP' radius of \citealt{stevens14}, also referred to as the `inherent' properties in \citealt{stevens19a}) in this figure and throughout this Letter.
We also exclude any galaxy with a dark-matter fraction below 5\% to conservatively remove non-cosmological objects \citep[cf.~the criteria in][]{nelson19b}.
This totals 2479 galaxies from TNG50.
Other than a small systematic increase in \Htwo~fraction for TNG50, the two simulations are well aligned.
There is generally good agreement between global \HI~and \Htwo~gas fractions in TNG100 and low-redshift observations \citep[e.g., ALFALFA, xGASS, and xCOLD GASS;][]{diemer19, stevens19b}, especially after mock-observing TNG galaxies to survey specifications. 
For the purposes of this paper, we infer by extension that TNG50 sufficiently agrees with observations. We note that in Figure \ref{fig:HIH2frac} the TNG50 cluster sample has higher gas fractions than their VERTICO counterparts at fixed stellar mass, particularly where $m_* \! \leq \! 10^{10}\,{\rm M}_\odot$. Due to the small number statistics in this regime, we caution against over interpretation, yet it is possible that this is due to a  systematic difference between the simulated and observed samples. For example, environmental processes in TNG50 may not be regulating gas content in lower stellar mass simulated galaxies to the same extent as for the observed galaxies.

We create mock surface density maps of \HI, \Htwo, and stellar mass for each TNG galaxy in our sample. 
These maps are matched to the specifications of the HERACLES and VERTICO data by convolving each map at the simulation's native resolution with a Gaussian kernel with a full width at half maximum of 1.2\,kpc, and then resampling with square pixels of length 0.65\,kpc. This physical scale matches the resolution of the observational data.
Each map is made face-on using the angular-momentum vector of all neutral gas (any gas that is not ionized) inside the `BaryMP' radius \citep{stevens14}, under the assumption that the observations have been correctly deprojected.

We separate TNG50 galaxies into two sub-samples after applying the above cuts.
The `field' sample, intended to be comparable to HERACLES, is selected to contain only central galaxies (those in the most massive {\sc subfind} subhalo) in haloes with virial mass $M_{\rm 200c} \! \leq \! 10^{12.2}\,{\rm M}_\odot$.
The field sample totals 1739 galaxies with an average number of 1514 pixels with $\Sigma_* \! > \! 1\,{\rm M}_{\odot}\,{\rm pc}^{-2}$ per galaxy.
The `cluster' sample, comparable to VERTICO, contains only satellites (galaxies that are not centrals) in the simulation's two clusters%
\footnote{Images of these clusters (and much more) can be found at \url{https://www.tng-project.org/media/}.
We could obtain many more Virgo-mass clusters using TNG100 or TNG300, but the main advantage of TNG50 is its superior resolution.
For a pixel of size $(650\,{\rm pc})^2$, a surface density of $1\,{\rm M}_\odot\,{\rm pc}^{-2}$\,---\,which we want to resolve down to\,---\,contains the equivalent mass of $\sim$5 gas elements in TNG50.
In TNG100, this would be less than half a gas element.
We therefore stick to TNG50 to minimise the potential for numerical effects to skew our results.}
with $M_{\rm 200c} \! \gtrsim \! 10^{14}\,{\rm M}_\odot$.
The most massive cluster, with $M_{\rm 200c} \! = \! 1.8 \! \times \! 10^{14}\,{\rm M}_{\odot}$, is very similar in mass to Virgo ($1.4$--$4.2 \! \times \! 10^{14}\,{\rm M}_\odot$; see table 1 of \citealt{boselli18} and references therein).
The other TNG50 cluster is about half this mass ($M_{\rm 200c} \! = \! 9.4 \! \times \! 10^{13}\,{\rm M}_{\odot}$), but has a different dynamical state, and has already been justified as a Virgo analogue by \citet{joshi21}.
The cluster sample totals 41 galaxies with an average of 1725 pixels with $\Sigma_* \! > \! 1\,{\rm M}_{\odot}\,{\rm pc}^{-2}$ per galaxy.
The higher pixel count per galaxy in the cluster sample reflects its higher average stellar mass (cf.~the distribution of TNG50 points in Fig.~\ref{fig:HIH2frac}).


\section{Results}

We explore how the resolved gas--stellar surface density scaling relations of galaxies are affected by environment, which we do in two steps.
First, we compare sample-averaged scaling relations between the field and cluster samples in TNG50 to those of HERACLES and VERTICO.
Second, we present the resolved relations of individual cluster galaxies, assessing how the relations change with global \HI~deficiencies, and comparing analogous galaxies between TNG50 and VERTICO.

\subsection{Sample averages}

\begin{figure}[t]
\centering
\includegraphics[width=\columnwidth]{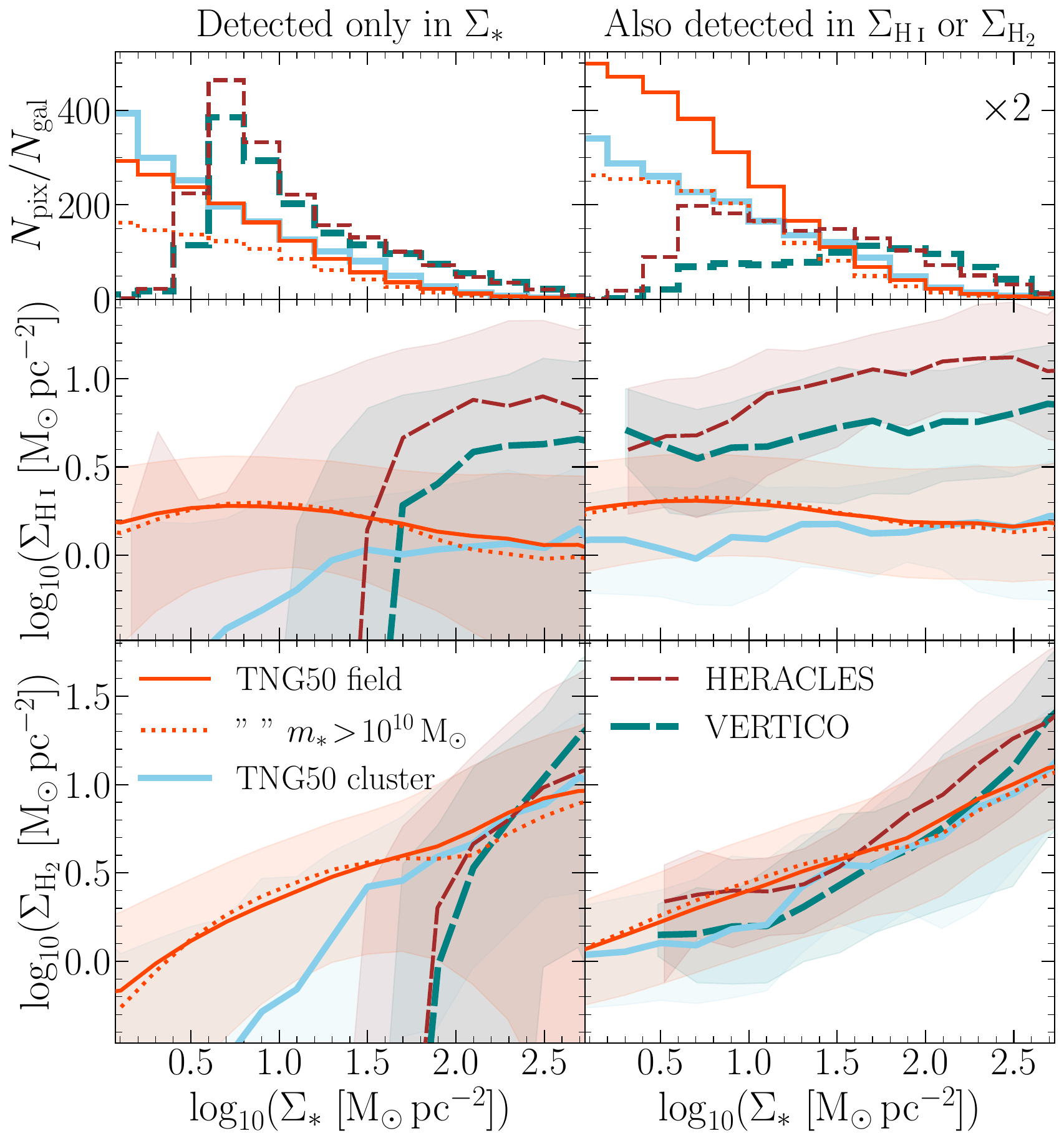}
\vspace{-0.2cm}\caption{Resolved scaling relations for TNG50 galaxies per their field and cluster samples (and a high-mass field sub-sample), compared respectively with HERACLES and VERTICO.
Lines are running medians in 0.2-dex bins of $\Sigma_*$.
Shaded regions cover the 16th to 84th percentiles (not shown for the high-mass field sub-sample).
The left column accounts for all pixels in both the observations (provided they were detected in stellar emission) and simulations, setting non-detections in either \SHI~or \SHtwo~in the observations to zero, which are accounted for in the percentiles.
The right column removes any non-detections in the observations by cutting out any pixels that would fall below the axes as plotted.
The lower boundary of each axis represents the 1st percentile of all gas-detected pixels (irrespective of whether the pixel is a detection in \Ss) across both observational surveys.
This boundary also represents the cut in gas surface density applied to TNG50 in the right-hand panels, as to emulate a detection threshold.
The top two panels show the one-dimensional histograms of \Ss~for pixels in each sample, normalised by the number of galaxies in that sample.
The $y$-axis in the top-right panel is stretched by a factor of two for clarity.
}
\label{fig:scalings}
\end{figure}

Our first aim is to see if the \emph{average} behavior of kpc-scale gas in TNG50 reflects that of reality in both the field and cluster environments.
In Fig.~\ref{fig:scalings}, we present the pixel-based relations of \SHI~and \SHtwo~as a function of \Ss.
The pixels for all galaxies in the TNG50 cluster sample are grouped together, as are those in the field sample.
The same grouping strategy applies for HERACLES and VERTICO.
To demonstrate that the preference of HERACLES galaxies to have $m_* \! \gtrsim \! 10^{10}\,{\rm M}_\odot$ (seen in Fig.~\ref{fig:HIH2frac}) does not affect our comparison, we add results to Fig.~\ref{fig:scalings} for a TNG50 field sub-sample where we have excluded galaxies with $m_* \! \leq \! 10^{10}\,{\rm M}_\odot$.
We exclude the resolved molecular Kennicutt--Schmidt relation (\SHtwo--$\Sigma_{\rm SFR}$) from Fig.~\ref{fig:scalings}, as we find that the relation is independent of environment to first order in TNG50.
Indeed, \citet{jimenez23} have also shown that this is true in observations.
Because of this, we also choose not to include the resolved star-forming main sequence (\Ss--$\Sigma_{\rm SFR}$), as it presents similar information to the resolved molecular gas main sequence in TNG50, as it does in observations (\Ss--\SHtwo, addressed further in \citealt{brown23}).
See \citet{motwani22} for further analysis on kpc-scale $\Sigma_{\rm SFR}$ in TNG50.

The left column of panels in Fig.~\ref{fig:scalings} accounts for all pixels with \Ss~detections, regardless of whether they were detected in \HI~or \Htwo.
For the purposes of calculating percentiles, gas non-detections are assumed to have zero mass.
The sharp downturn in the VERTICO and HERACLES medians from right to left is simply representative of the threshold surface densities needed for gas to be detected in those surveys.

From these panels, we can immediately identify that TNG50 quantitatively reproduces the correlation between \SHtwo~and \Ss, while the \SHI--\Ss~relation in the simulation is systematically offset to lower \SHI~values. VERTICO cluster galaxies are systematically suppressed in \SHI~by a factor of $\sim$2 at fixed \Ss~relative to HERACLES, but not significantly for \SHtwo.
This is in line with the findings of \citet{watts23}, even though we are including lower-mass galaxies than in that study.

By contrast to the observations, the \SHI~medians are not parallel for the field and cluster samples in TNG50, instead converging at high \Ss.
Both \SHI~and \SHtwo~become increasingly divergent between the TNG50 cluster and field samples at $\Sigma_* \! \lesssim \! 30\,{\rm M}_\odot\,{\rm pc}^{-2}$.
This average behavior is typical of disc truncation, which one would expect from processes like ram-pressure stripping that affect the cluster galaxies. 

TNG50 galaxies in field and cluster samples are systematically low in \SHI~relative to observations and exhibit little to no environmental dependence at high \Ss~(the centers of galaxies).
This central deficit in \HI~appears to be widespread in TNG50, a finding described in \citet{gebek23}, and also seen in TNG100 by \citet{diemer19} and in 
figures A1 and A2 of \citet{stevens19b}.
A similar outcome has also been found for the EAGLE simulation \citep{bahe16}.
Discussion in section 4.5.2 of \citet{diemer19} and section 4.3 of \citet{gebek23} describe how (some) central \HI~deficits in TNG are likely the result of high ionized fractions in the interstellar medium and AGN feedback removing gas from galaxy centers \citep[also see][]{stevens19a,stevens21}.
We also note that among all the post-processing methods used to decompose neutral gas in TNG into its atomic and molecular components, all have molecular fractions that asymptote to 1 at high densities, evidently leading to a relatively low saturation density for \HI~\citep[see][]{diemer18}.

There are also fewer pixels per galaxy at high \Ss~in TNG50 than observations, as seen in the top panels of Fig.~\ref{fig:scalings}.
This implies that galaxy centers are generally under-dense in the simulation.
Numerical heating of stellar particles plays a role here in artificially decreasing central stellar densities \citep[see][]{ludlow21}.
This same effect has a minimal influence on gas, suggesting pixels in TNG50 galaxies are leftward of where they should be in Fig.~\ref{fig:scalings}. The low gas densities, combined with this systematic underestimation of \Ss~in the centers of galaxies, is likely also the reason that we do not see an environmental influence on \SHI~at high values of \Ss~compared to observations.

It is important to assess what the role of pixels undetected in gas have on (i) the average behavior of the observations and (ii) the comparison with TNG described above.
To this end, we compare gas-detected pixels in both the observations and simulations in the right-hand panels of Fig.~\ref{fig:scalings}. 
We apply a lower limit for \SHI~and \SHtwo~equal to the minimum value in the respective axes in the figure to emulate a detection threshold for TNG.
While the median \Ss--\SHtwo~relations for \SHtwo~detections in TNG50 and observations are in reasonable agreement, the systematic deficit in TNG's \SHI~relative to observations is even more stark than before.
The latter suggests that TNG galaxies are already suppressed in \SHI~at high \Ss~before falling into the cluster.
There is minimal suppression in their central \SHI~thereafter, while the observations instead show a clear suppression.
However, because the TNG galaxies are selected to be star-forming, and only 41 of those meet our cluster criteria, this does \emph{not} automatically mean the correct qualitative effect of environment on kpc-scale gas is absent in the simulation, as we explore in the next subsection.

\subsection{Individual cluster galaxies}

\begin{figure*}[t]
\centering
\includegraphics[width=2.0\columnwidth]{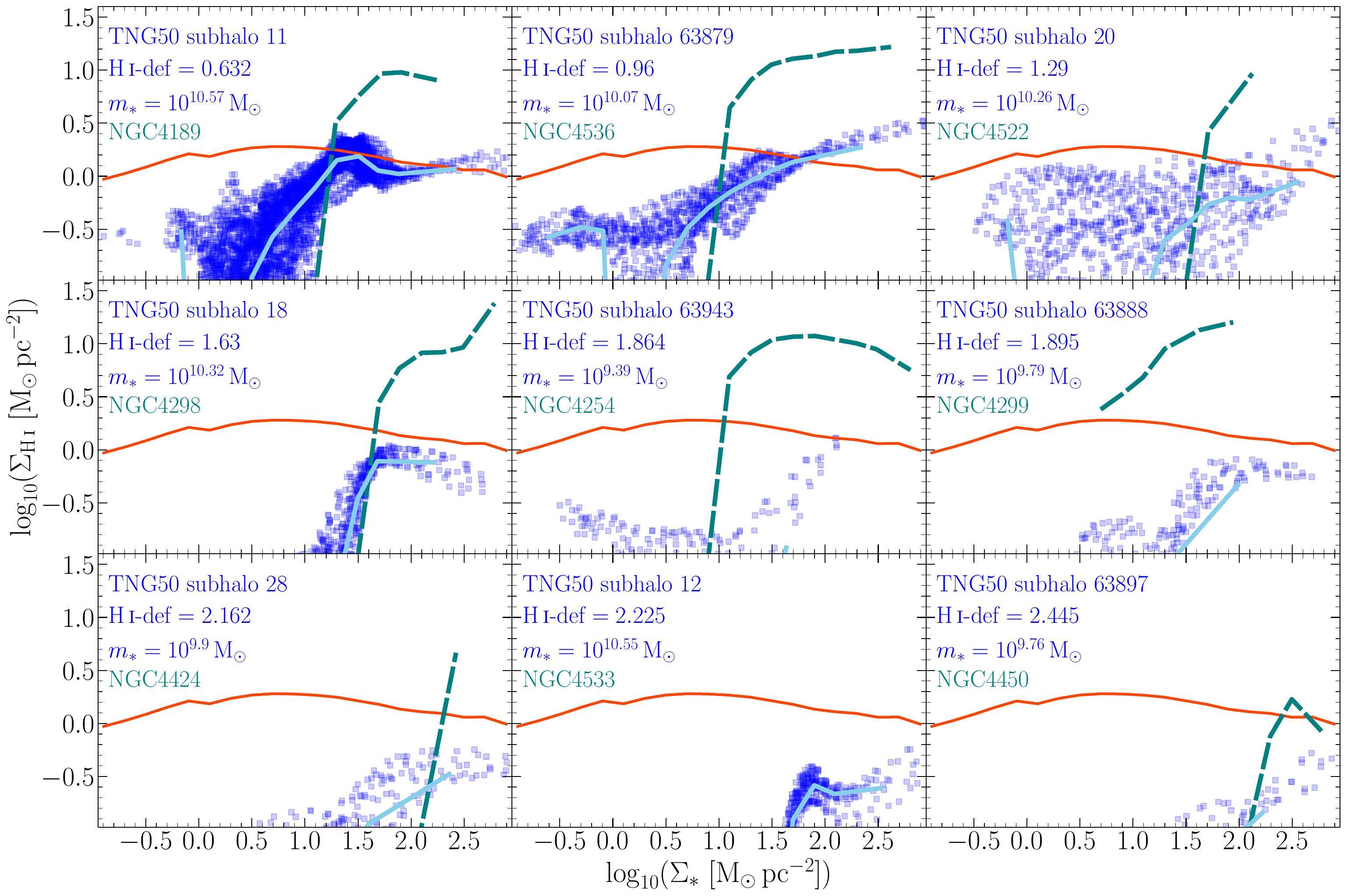}
\vspace{-0.2cm}\caption{Individual \HI~sequences for the nine most \HI-deficient galaxies among VERTICO analogues in TNG50.
Points are pixels from the TNG50 galaxies, with thick solid lines the running median of those points.
The thin, solid, red line that repeats in each panel is the median for the TNG50 field sample.
Each dashed line is the running median for the VERTICO galaxy that the TNG50 galaxy is matched to, based on its stellar mass and distance from the star-forming main sequence. 
NGC4533 lacks any detected resolved \HI.
The lower bound of the $y$-axis in each panel is $\sim$0.5\,dex lower than what is detected in VERTICO galaxies.
}
\label{fig:HI}
\end{figure*}

\begin{figure*}[t]
\centering
\includegraphics[width=2.0\columnwidth]{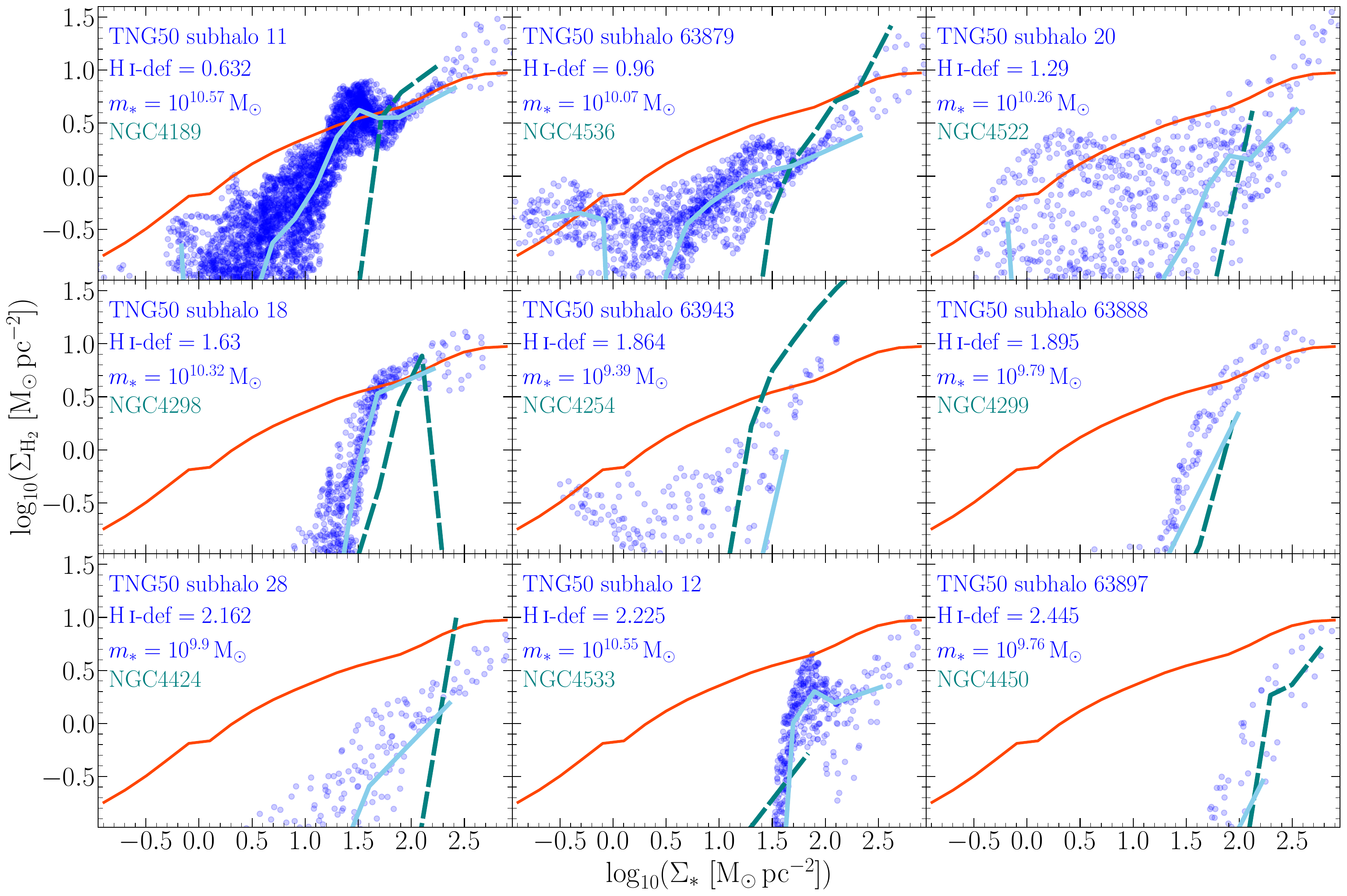}
\vspace{-0.2cm}\caption{Same as Fig.~\ref{fig:HI} but for the resolved molecular-gas main sequence.
}
\label{fig:H2}
\end{figure*}

The ensemble relations above demonstrate the systematic underestimation of \SHI~at fixed stellar density in TNG50 with respect to the observations.
But ensemble relations are merely the superposition of individual relations.
Thus, to establish if TNG50 can reproduce the
resolved effects of environmental mechanisms on individual galaxies' gas content reported by \citet{watts23}, we need to explore their \emph{individual} \Ss--\SHI~and \Ss--\SHtwo~sequences.

In a similar fashion to \citet{watts23}, we rank order our TNG50 cluster sample by \HI~deficiency (\HI-def for short), and show their individual \Ss--\SHI~and \Ss--\SHtwo~sequences in Figs \ref{fig:HI} and \ref{fig:H2}, respectively.
\HI~deficiency is often used as a proxy for how processed a galaxy has been by its environment, provided \HI-def $\! > \! 0.3$.
Here, \HI-def is defined as the logarithmic deviation in expected \HI~mass at fixed stellar mass, relative to the median of TNG50 field galaxies (shown in Fig.~\ref{fig:HIH2frac}).%
\footnote{This is a different definition of \HI-def to that used in \citet{watts23}, but it serves the same function.}
Each pixel for each TNG50 galaxy in Figs \ref{fig:HI} and \ref{fig:H2} is shown along with their running median.
The running median for the whole TNG50 field sample (a `control') is shown along with one analogue VERTICO galaxy, matched according to its stellar mass and distance from the (integrated) star-forming main sequence.
For the real galaxies, we use the main sequence of \citet{leroy19}, which we write in terms of specific star formation rate:
\begin{equation}
\log_{10}\! \left( \frac{{\rm sSFR}_{\rm MS}^{\rm L19}}{{\rm yr}^{-1}} \right) = -0.346\, \log_{10} \! \left( \frac{m_*}{{\rm M}_\odot} \right) - 6.652\,.
\end{equation}
For TNG galaxies, we iteratively perform a least-squares straight-line fit to $\log_{10}(m_*)$--$\log_{10}({\rm sSFR})$, removing outliers of $>\!3\sigma$ each time to ensure we do not accidentally include quenched galaxies, until the fit converges.
The resulting fit is
\begin{equation}
\log_{10}\! \left( \frac{{\rm sSFR}_{\rm MS}^{\rm TNG}}{{\rm yr}^{-1}} \right) = -0.159\, \log_{10} \! \left( \frac{m_*}{{\rm M}_\odot} \right) - 8.427\,.
\end{equation}
The vector length between the TNG50 and VERTICO matched pairs is always shorter than 0.3\,dex in this parameter space.
Although \HI~deficiency and distance from the main sequence are physically correlated, we note that this method does \emph{not} mean that the \HI~deficiencies of the paired VERTICO and TNG50 galaxies are the same. 
There are 13 galaxies in the TNG50 cluster sample that both have \HI-def $\! > \! 0.3$\,dex and have VERTICO analogues, of which the nine with the highest \HI-def are shown in Figs \ref{fig:HI} and \ref{fig:H2}.

Fig.~\ref{fig:HI} demonstrates that when examining the most environmentally affected TNG galaxies, the simulation \emph{does} reproduce the \emph{qualitative} results of VERTICO per \citet{watts23}; i.e.~that (i) gas discs are truncated to higher \Ss~for higher \HI-def and (ii) central \SHI~decreases for high \HI-def.
This did not appear to be the case in Fig.~\ref{fig:scalings} because 28 of the 41 galaxies in the TNG50 cluster sample are either \HI-normal, i.e.~they have \HI-def $\! < \! 0.3$, and/or have no (unique) analogue in VERTICO.

Fig.~\ref{fig:H2} shows the \Ss--\SHtwo~sequences for the same galaxies as in Fig.~\ref{fig:HI}.
We see truncation occurring at the same \Ss~in both \SHI~and \SHtwo, and we see a suppression in central \SHtwo~for the most \HI-deficient galaxies.
These results are again qualitatively in line with that reported for VERTICO in \citet{watts23}.

The fact that the majority of galaxies in the TNG50 cluster sample are unaffected by their environment (i.e.,  28 of the 41 galaxies are \HI-normal, a much larger fraction than in the observations) is because many TNG galaxies that are strongly affected by a cluster environment belong to a subhalo that is devoid of gas cells entirely,%
\footnote{Gas cells in {\sc arepo} are volume-filling, meaning gas is always present everywhere. But that does not mean gas is specifically associated with substructure or a galaxy.}
especially at low stellar masses (see section 5.2 of \citealt{diemer19}; figure 9 of \citealt{stevens21}).
Naturally, such galaxies, where the environmental influence on gas content is greatest, cannot be included in this work.
While the improved resolution of TNG50 relative to earlier simulations in the suite certainly combats this issue, numerical simulations are always limited by discretization.
Only with enough dense gas elements are the hydrodynamical forces of environment reliable in the simulation.
In essence, this means there is a narrow window of opportunity to catch simulated galaxies experiencing the onset of environmental effects.
No analogous limitation exists for observed galaxies.
This might explain why the TNG50 cluster sample is small in number, despite having two clusters of comparable mass to Virgo.
In future work, this issue can be mitigated by using more snapshots to catch the moment of interest for each infalling satellite galaxy.


\section{Summary}

TNG50 \emph{quantitatively} reproduces the \Ss--\SHtwo~relation found in observations in both cluster and field samples. The \Ss--\SHI~relation, on the other hand, is found to be endemically gas-poor at fixed \Ss~with respect to the observations.
In addition, we find that the kpc-scale effects of a Virgo-like environment on satellite galaxies' \HI~and \Htwo~gas content is \emph{qualitatively} recovered by TNG50 at $z\!=\!0$.
Gas discs are not only truncated more the more they have been affected by their environment, but their central gas densities are also relatively suppressed.
However, this effect is likely quantitatively weaker than in reality, because
the central gas surface densities of TNG50 galaxies, particularly in \HI, are systematically low relative to observations, irrespective of environment.

With this baseline performance of TNG50 established, it opens the door to using the TNG simulations to show how the resolved gas scaling relations of galaxies change after infall into a cluster or otherwise dense environment.
Such an experiment is crucial to reinforce the theoretical interpretation of the empirical results of VERTICO, which must rely on conjecture in describing cluster galaxies' histories.


\section*{Acknowledgments}
Parts of this research were funded by a Research Collaboration Award from The University of Western Australia.
ARHS is a grateful recipient of the Jim Buckee Fellowship at ICRAR-UWA.
TB acknowledges support from the National Research Council of Canada via the Plaskett Fellowship of the Dominion Astrophysical Observatory.
VV acknowledges support from the scholarship ANID-FULBRIGHT BIO 2016 - 56160020 and funding from NRAO Student Observing Support (SOS) - SOSPADA-015.
DN acknowledges funding from the Deutsche Forschungsgemeinschaft (DFG) through an Emmy Noether Research Group (grant number NE 2441/1-1). YB acknowledges funding from the Dutch Research Organisation (NWO) through Veni grant number 639.041.751 and financial support from the Swiss National Science Foundation (SNSF) under project 200021\_213076.

This paper makes use of the following ALMA data: 
\begin{itemize}
    \item ADS/JAO.ALMA\href{https://almascience.nrao.edu/asax/?result_view=observation&projectCode=\%222019.1.00763.L\%22}{\#2019.1.00763.L} 
    \item ADS/JAO.ALMA\href{https://almascience.nrao.edu/asax/?result_view=observation&projectCode=\%222017.1.00886.L\%22}{\#2017.1.00886.L} 
    \item ADS/JAO.ALMA\href{https://almascience.nrao.edu/asax/?result_view=observation&projectCode=\%222016.1.00912.S\%22}{\#2016.1.00912.S} 
    \item ADS/JAO.ALMA\href{https://almascience.nrao.edu/asax/?result_view=observation&projectCode=\%222015.1.00956.S\%22}{\#2015.1.00956.S}
\end{itemize}
ALMA is a partnership of ESO (representing its member states), NSF (USA) and NINS (Japan), together with NRC (Canada), MOST and ASIAA (Taiwan), and KASI (Republic of Korea), in cooperation with the Republic of Chile. The Joint ALMA Observatory is operated by ESO, AUI$/$NRAO and NAOJ. The National Radio Astronomy Observatory is a facility of the National Science Foundation operated under cooperative agreement by Associated Universities, Inc.

This research made use of data provided by the NASA/IPAC Infrared Science Archive, which is funded by the National Aeronautics and Space Administration and operated by the California Institute of Technology.

The authors acknowledge the use of the Canadian Advanced Network for Astronomy Research (CANFAR) Science Platform. Our work used the facilities of the Canadian Astronomy Data Center, operated by the National Research Council of Canada with the support of the Canadian Space Agency, and CANFAR, a consortium that serves the data-intensive storage, access, and processing needs of university groups and centers engaged in astronomy research \citep{gaudet10}.

\vspace{5mm}
\facilities{ALMA, GALEX, Sloan, WISE}

\bibliography{references}{}
\bibliographystyle{aasjournal}

\end{document}